\documentclass{astlb}
\usepackage{graphicx}
\usepackage{natbib}
\usepackage{color}

\usepackage[hyperfootnotes=false]{hyperref}

\usepackage{multirow}

\definecolor{darkblue}{rgb}{0,0,0.9}

\def\smfigure#1#2#3{
  \begin{minipage}{1.0\columnwidth}
    \begin{minipage}{0.049\columnwidth}
      \rotatebox{90}{\small\phantom{0000}#3}
    \end{minipage}
    \begin{minipage}{0.9\columnwidth}
      \includegraphics[bb=40 188 556 678,width=0.97\columnwidth]{#1}
      \centerline{\small #2}
    \end{minipage}

    \vskip 3pt
    ~
  \end{minipage}
}

\def\neff{\ensuremath{N_{\mathrm{eff}}}}

\begin{document}

\journalinfo{2013}{39}{0}{1}[0]

\title{Possible indication for non-zero neutrino mass and additional
  neutrino species from cosmological observations}

\author{R.~A.~Burenin\email{rodion@hea.iki.rssi.ru}\address{1}
  \addresstext{1}{Space Research Institute of RAS (IKI), Moscow}
}

\shortauthor{R. A. Burenin}

\shorttitle{Indication for non-zero neutrino mass and additional
  neutrino species}

\submitted{December 21, 2012}

\begin{abstract}
  The constraints on total neutrino mass and effective number of
  neutrino species based on CMB anisotropy power spectrum, Hubble
  constant, baryon acoustic oscillations and galaxy cluster mass
  function data are presented. It is shown that the discrepancies
  between various cosmological data in Hubble constant and density
  fluctuation amplitude, measured in standard $\Lambda$CDM
  cosmological model, can be eliminated if more than standard
  effective number of neutrino species and non-zero total neutrino
  mass are considered. This extension of $\Lambda$CDM model appears to
  be $\approx3\sigma$ significant when all cosmological data are
  used. The model with approximately one additional neutrino type,
  $\neff\approx 4$, and with non-zero total neutrino mass, $\Sigma
  m_\nu\approx 0.5$~eV, provide the best fit to the data. In the model
  with only one massive neutrino the upper limits on neutrino mass are
  slightly relaxed. It is shown that these deviations from
  $\Lambda$CDM model appear mainly due to the usage of recent data on
  the observations of baryon acoustic oscillations. Larger than
  standard number of neutrino species is measured mainly due to the
  comparison of the BAO data with direct measurements of Hubble
  constant, which was already noticed earlier. As it is shown below,
  the data on galaxy cluster mass function in this case give the
  measurement of non-zero neutrino mass.
  
  \keywords{cosmology, cosmological parameters, effective number of
    neutrino species, total neutrino mass}

\end{abstract}

\section{Introduction}
\label{sec:intro} 

The existence of neutrino with masses in the eV mass range would
produce a suppression of density fluctuations on scales below the
horizon when these neutrinos become non-relativistic
\citep[e.g.,][]{hu98}. The amount of this suppression can be measured
from the comparison of density fluctuations amplitude measured at
early epoch from CMB anisotropy power spectrum and at recent epoch,
from, e.g., galaxy cluster mass function data, which finally could
give the measurement of total neutrino mass.

From cosmological observations the number of neutrino species could
also be measured. Any type of neutrino, which was thermalized in early
Universe, make its own contribution into the energy density of
relativistic matter before equipartition. The change of this energy
density result in a change in Universe expansion rate and therefore in
changes of the size of sound horizon and photon diffusion scale (Silk
damping scale).

All these quantities can be measured using the data from various
cosmological observations. For example, the information on density
fluctuation amplitude at early epoch, size of sound horizon, Silk
damping scale is contained in CMB anisotropy power spectrum which is
now accurately measured in various experiments, such as \emph{WMAP}
\citep[e.g.,][]{larson11}, South Pole Telescope
\citep[]{keisler11,story12}, Atacama cosmology telescope
\citep{dunkley11}, in very near future the results of CMB measurements
obtained with Planck space observatory should also be published
\citep{planck11}.

The linear density fluctuations amplitude at recent epoch, which is
usually described using $\sigma_8$ parameter, may be measured by means
of various methods. One of the most accurate measurement of this
quantity comes from the data on galaxy cluster mass function
\citep[see, e.g.,][]{av09a,av09b,mantz10a,benson11,bv12}.

The information on the size of sound horizon is contained also in the
baryon acoustic oscillations data \citep[BAO, see,
e.g.,][]{percival10}. In addition, the direct measurements of Hubble
constant \citep[e.g.,][]{riess11} should also be used in order to
eliminate degeneracies between various cosmological parameters. The
combination of all the data discussed above allowed to obtain
constraints on total neutrino mass and number of neutrino species
\citep[e.g.,][]{av09b,mantz10b,keisler11,benson11,bv12}.

In addition to the data used in these works, substantially improved
new data on BAO observations were published recently
\citep{blake11,beutler11,padmanabhan12,anderson12}. It was noticed
that the distance scale measured using these data in assumption of
standard $\Lambda$CDM model turns out to be in some tension with the
results of direct Hubble constant measurements and that this
discrepancy can be eliminated with the assumption of larger than
standard effective neutrino species \citep{mehta12,anderson12}. With
new additional data on Hubble constant measurements taken in account
\citep{freedman12}, this discrepancy have increased.

As it is shown below, when these new BAO data are used, there is also
a discrepancy in $\sigma_8$ measurements inferred from the data on CMB
anisotropy power spectrum in assumption of standard $\Lambda$CDM model
and obtained using the galaxy cluster mass function data from
\cite{av09a,av09b}. This discrepancy can be interpreted as a
suppression of density fluctuations due to the existence of neutrinos
with non-zero total mass. It is shown that the best agreement with all
cosmological data discussed above is achieved in model with non-zero
neutrino mass and additional neutrino species.

The cosmological parameters constraints were calculated from the
simulations of Monte-Carlo Markov Chains, which were done using
\texttt{CosmoMC} software \citep{lewis02}, version of Jan.\ 2012. In
all Figures below the contours at 68\% and 95\% confidence levels are
shown. All numerical values of confidence intervals are given at 68\%
confidence level.

\section{Cosmological data}
\label{sec:data}

\subsection{Galaxy clusters}
\label{sec:cl}

For our work the data on galaxy cluster mass function measurements
were taken without any changes from \cite{av09a,av09b}. In this work a
sample of 86 massive galaxy clusters with masses measured with about
10\% accuracy using \emph{Chandra} observations \citep{av09a} was
used. Distant clusters, located at $z\approx0.4$--$0.9$, were selected
from 400d X-ray galaxy cluster survey, based on ROSAT pointing data
\citep{400d}. Clusters in local Universe were selected using ROSAT all
sky survey \citep[see details in][]{av09a}.

Likelihood functions for this cosmological dataset are available at
WWW\footnote{http://hea.iki.rssi.ru/400d/cosm/} (see also details in
\citealt{bv12}). Systematic uncertainties are not included in these
likelihood functions. These uncertainties are discussed in detail in
\cite{av09b} and can be taken in account separately
\citep{av09b,bv12}, which is done below for all measurements where
these data are used. This dataset is designated below as \emph{CL}.

\subsection{CMB power spectrum}
\label{sec:cmb}

For our work we used the data of 7-year observations of \emph{WMAP}
observatory \citep{larson11,komatsu11}. Likelihoods were calculated
using the software taken from
archive\footnote{http://lambda.gsfc.nasa.gov/}, version
\emph{4.1}. Also we used the data at small angular scale CMB
anisotropy obtained with South Pole Telescope
\citep[\emph{SPT},][]{keisler11}. When using these data, the
contribution of Sunyaev-Zeldovich effect and also the contributions of
poisson and clustered sources were taken into account according to the
prescription of \S 4.1 in \cite{keisler11}. These data taken together
are designated below as \emph{CMB}.

\subsection{Hubble constant}
\label{sec:h0}

In our work we used the Hubble constant measurement obtained using
improved calibration of supernovae type Ia absolute magnitudes,
$H_0=73.8\pm2.4$~km~s$^{-1}$Mpc$^{-1}$ \citep{riess11}. The error here
includes both statistical and systematic uncertainties. In addition to
this measurement, the results of Carnegie Hubble Project were
published recently, where the new Cepheid distance scale calibration,
obtained using the data of Spitzer Space Telescope, was applied to the
Hubble Space Telescope Key Project data and the measurement
$H_0=74.3\pm1.5\mbox{(stat.)}\pm2.1\mbox{(sys.)}$~km~s$^{-1}$Mpc$^{-1}$
was obtained \citep{freedman12}.

As compared to \cite{riess11}, in this work the distance to maser
galaxy \emph{NGC4258} was not used. Also, this work is based on a new,
independent determination of the distance to LMC, therefore, the
calibration of Cepheid period--luminosity relation should be
considered as independent one. For distance measurements different
data were also used (see details in \citealt{riess11} and
\citealt{freedman12}). Therefore, apparently, the Hubble constant
measurements presented in these two works should also be considered as
independent ones.

These two measurements combined in assumption of their independence and
gaussian errors give the value
$H_0=74.1\pm1.8$~km~s$^{-1}$Mpc$^{-1}$. This measurement is designated
below as \emph{$H_0$}. We note, however, that new Hubble constant
measurement from \cite{freedman12} do not produce strong changes in
the constraints presented below. The constraints obtained using $H_0$
measurement from \cite{riess11} only are also given below for
reference.

\subsection{Baryon acoustic oscillations}
\label{sec:bao}

The results of new, considerably improved measurements of baryon
acoustic oscillations made using the data of large spectroscopic
surveys of galaxies, published recently, are also used in our
work. They include the reprocessed data of SDSS Data Release 7
\citep{padmanabhan12}, the data of \emph{WiggleZ} \citep{blake11} and
\emph{6dF} \citep{beutler11} surveys, as well as the measurements made
using the data of SDSS Data Release 9 \citep[\emph{BOSS} survey,
\emph{CMASS} sample,][]{anderson12}. All these data taken together are
designated below as \emph{BAO}.

\begin{figure*}
  \centering
  \begin{minipage}{0.48\linewidth}
    \smfigure{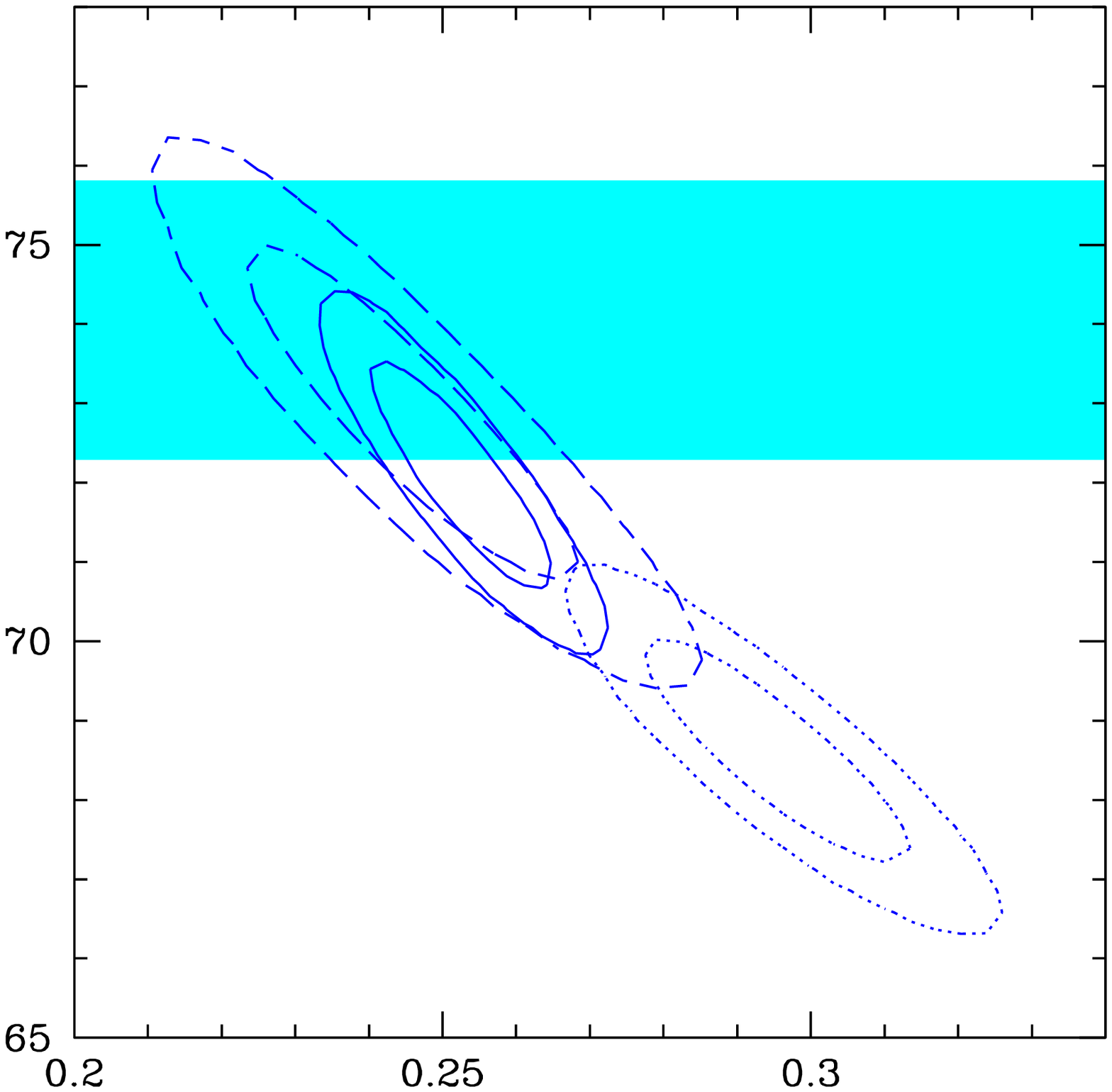}{$\Omega_m$}{$H_0$, km s$^{-1}$ Mpc$^{-1}$}
    \vskip -7.9cm \hskip 1.25cm {\footnotesize $\Lambda$CDM}
    \vskip 7.04cm ~
    \vskip -7.1cm \hskip 1.6cm {\footnotesize \emph{CMB}+$H_0$}
    \vskip 6.24cm ~
    \vskip -6.8cm \hskip 6.9cm {\footnotesize $H_0$}
    \vskip 5.94cm ~
    \vskip -3.7cm \hskip 2.4cm {\footnotesize \emph{CMB}+\emph{CL}}
    \vskip 2.84cm ~
    \vskip -2.cm \hskip 4.3cm {\footnotesize \emph{CMB}+\emph{BAO}}
    \vskip 1.14cm ~
  \end{minipage}
  \begin{minipage}{0.48\linewidth}
    \smfigure{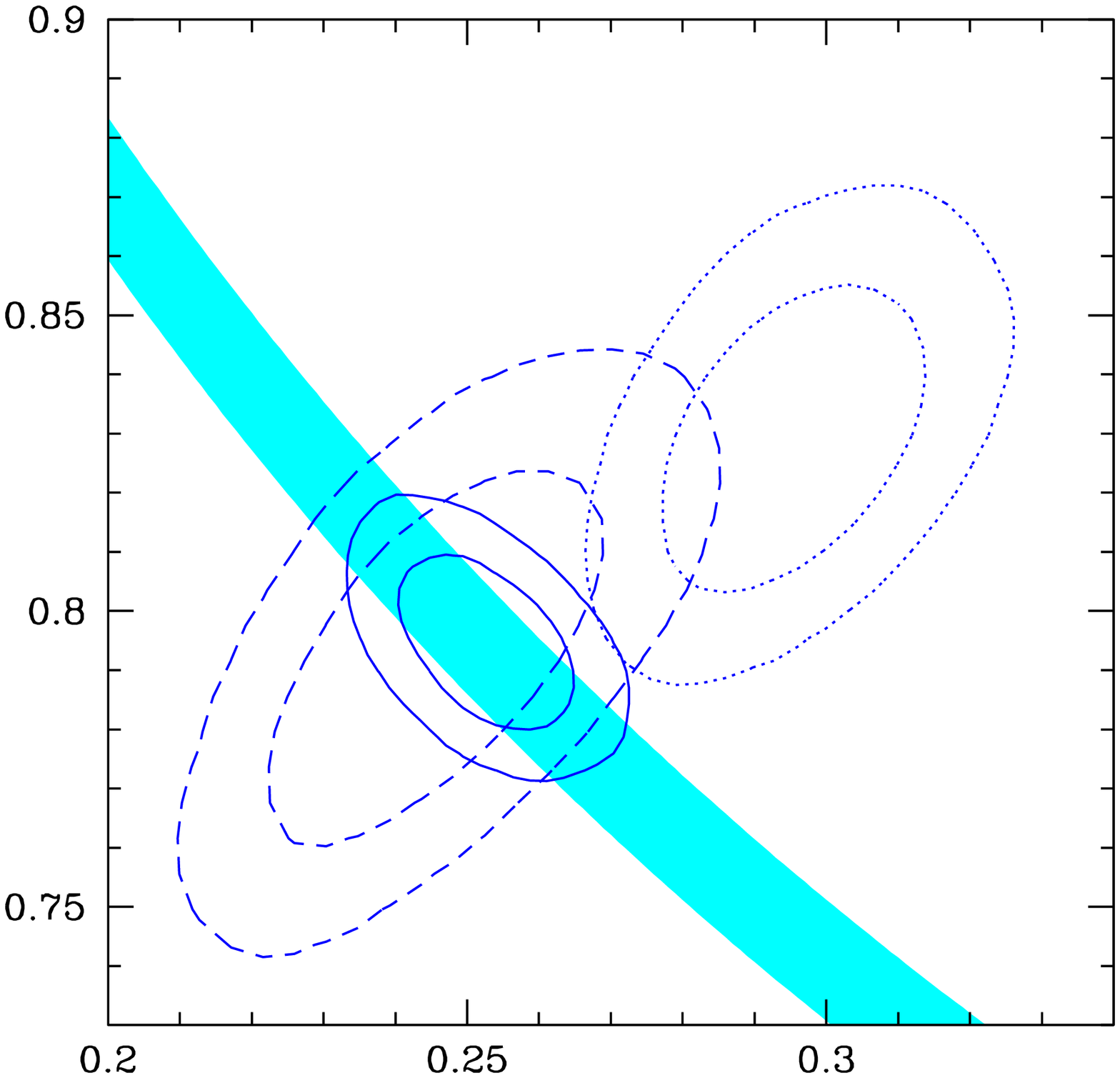}{$\Omega_m$}{$\sigma_8$}
    \vskip -7.9cm \hskip 1.25cm {\footnotesize $\Lambda$CDM}
    \vskip 7.04cm ~
    \vskip -6.8cm \hskip 5.5cm {\footnotesize \emph{CMB}+\emph{BAO}}
    \vskip 5.94cm ~
    \vskip -2.9cm \hskip 4.5cm {\footnotesize \emph{CMB}+\emph{CL}}
    \vskip 2.04cm ~
    \vskip -5.7cm \hskip 2.6cm {\footnotesize \emph{CMB}+\emph{$H_0$}}
    \vskip 4.84cm ~
    \vskip -1.6cm \hskip 6.5cm {\footnotesize \emph{CL}}
    \vskip 0.74cm ~
  \end{minipage}
  \caption{The constraints on $\Omega_m$, $\sigma_8$ and $H_0$ in
    $\Lambda$CDM model, obtained using various cosmological
    datasets. The constraints from \emph{CMB}+\emph{CL} data are shown
    with solid lines, from \emph{CMB}+$H_0$ data --- with dashed
    lines, from \emph{CMB}+\emph{BAO} data --- with dotted
    lines. Shaded regions show model independent constraints (at
    $1\sigma$ level) from direct measurements of Hubble constant
    (left) and for the combination of $\sigma_8$ and $\Omega_m$ from
    galaxy cluster mass function (right).}
  \label{fig:lcdmoms8h0}
\end{figure*}

\begin{figure*}
  \centering
  \begin{minipage}{0.48\linewidth}
    \smfigure{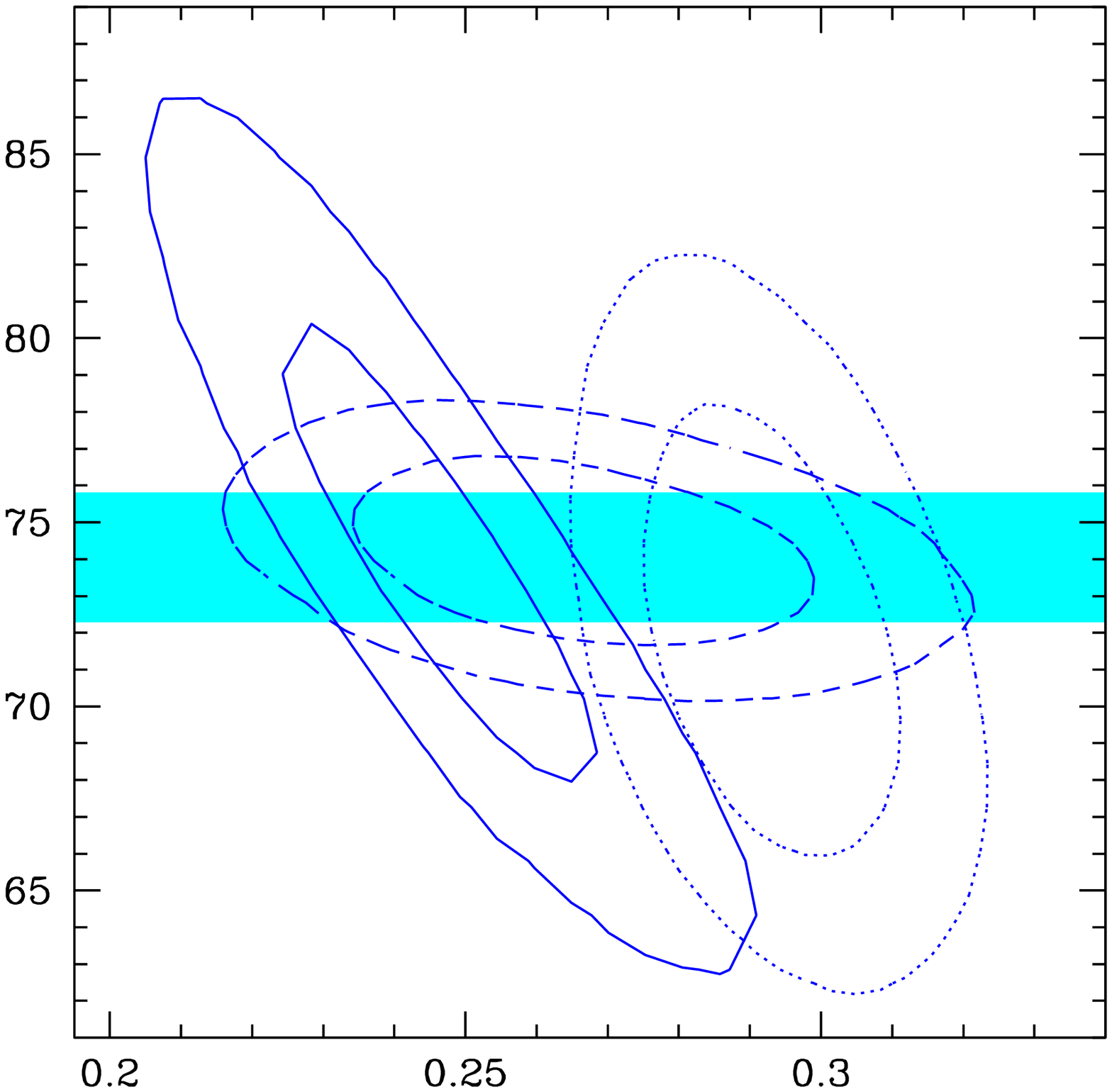}{$\Omega_m$}{$H_0$, km s$^{-1}$ Mpc$^{-1}$}
    \vskip -7.9cm \hskip 1.25cm {\footnotesize $\Lambda$CDM+\neff}
    \vskip 7.04cm ~
    \vskip -5.4cm \hskip 3.9cm {\footnotesize \emph{CMB}+$H_0$}
    \vskip 4.54cm ~
    \vskip -4.85cm \hskip 6.8cm {\footnotesize $H_0$}
    \vskip 3.99cm ~
    \vskip -7.cm \hskip 2.3cm {\footnotesize \emph{CMB}+\emph{CL}}
    \vskip 6.14cm ~
    \vskip -6.3cm \hskip 4.8cm {\footnotesize \emph{CMB}+\emph{BAO}}
    \vskip 5.44cm ~
  \end{minipage}
  \begin{minipage}{0.48\linewidth}
    \smfigure{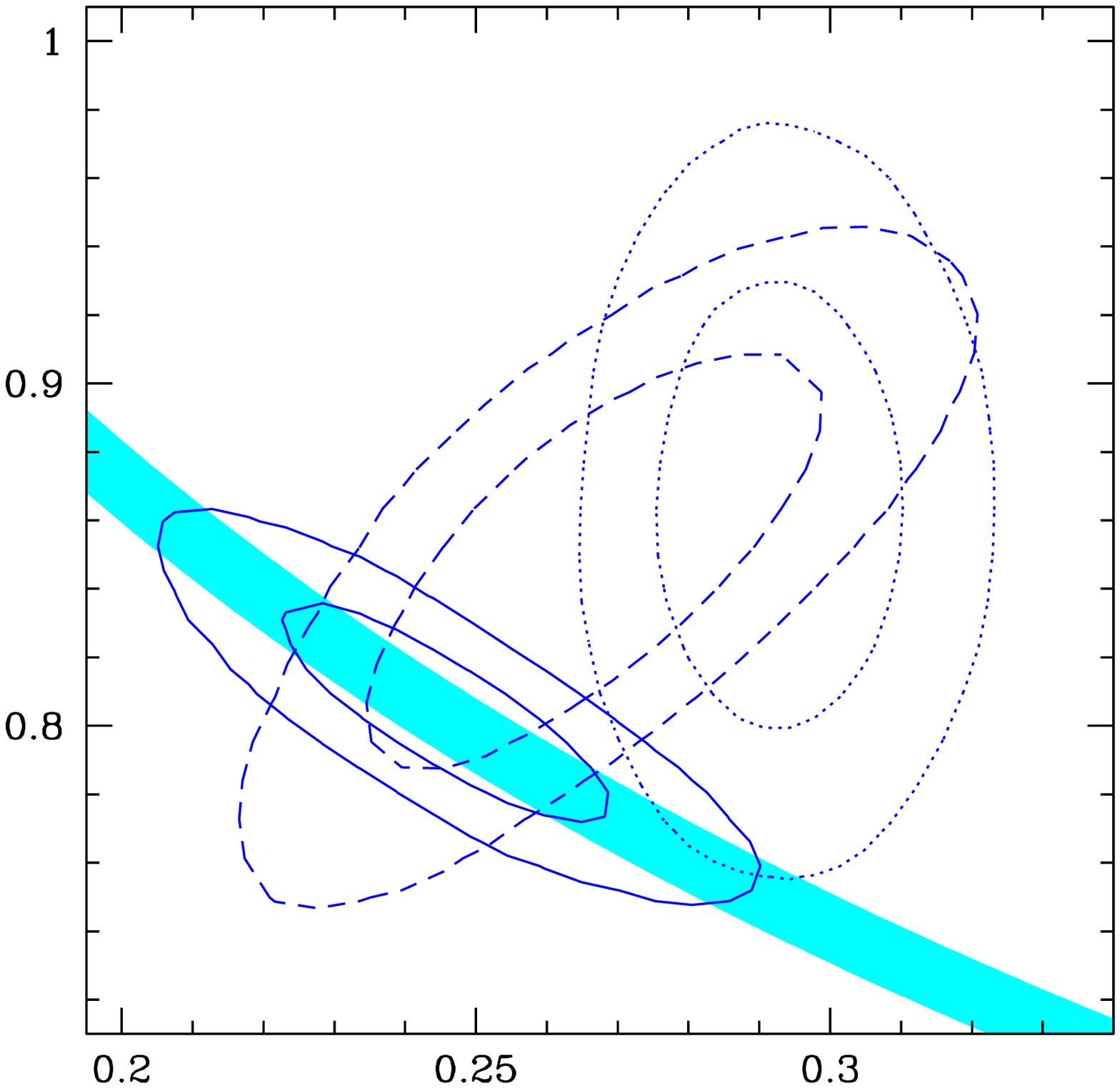}{$\Omega_m$}{$\sigma_8$}
    \vskip -7.9cm \hskip 1.25cm {\footnotesize $\Lambda$CDM+\neff}
    \vskip 7.04cm ~
    \vskip -7.1cm \hskip 5.cm {\footnotesize \emph{CMB}+\emph{BAO}}
    \vskip 6.24cm ~
    \vskip -1.8cm \hskip 4.cm {\footnotesize \emph{CMB}+\emph{CL}}
    \vskip 0.94cm ~
    \vskip -5.6cm \hskip 2.6cm {\footnotesize \emph{CMB}+$H_0$}
    \vskip 4.74cm ~
    \vskip -1.8cm \hskip 6.9cm {\footnotesize \emph{CL}}
    \vskip 0.94cm ~
  \end{minipage}
  \caption{The constraints on $\Omega_m$, $\sigma_8$ and $H_0$ in
    $\Lambda$CDM+\neff\ model, obtained using various cosmological
    datasets. Contours are the same as in Fig.~\ref{fig:lcdmoms8h0}.}
  \label{fig:lcdmoms8h0nnu}
\end{figure*}

\begin{figure*}
  \centering
  \begin{minipage}{0.48\linewidth}
    \smfigure{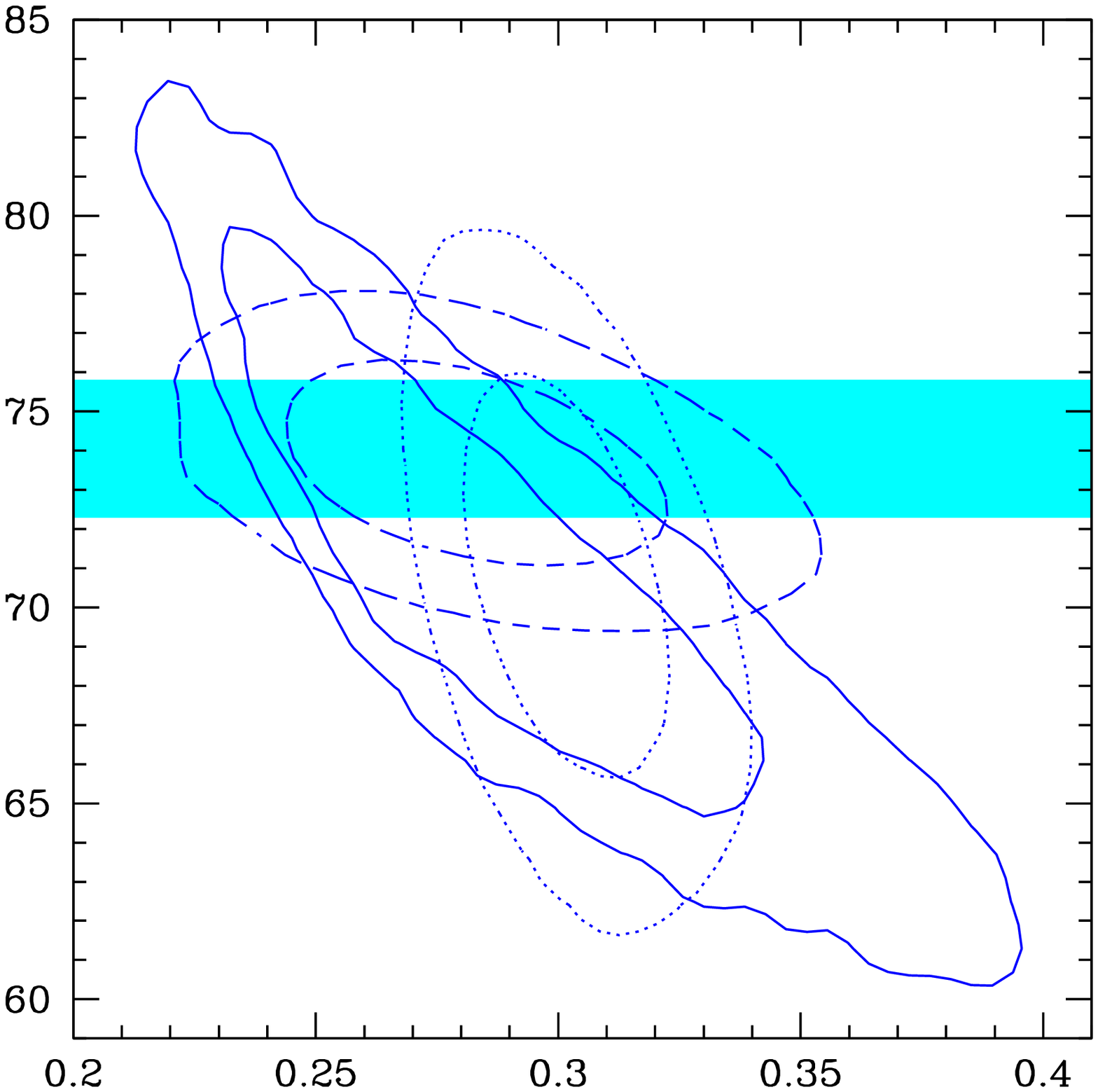}{$\Omega_m$}{$H_0$, km s$^{-1}$ Mpc$^{-1}$}
    \vskip -7.9cm \hskip 1.25cm {\footnotesize
      $\Lambda$CDM+\neff+$\Sigma m_\nu$}
    \vskip 7.04cm ~
    \vskip -4.1cm \hskip 5.9cm {\footnotesize \emph{CMB}+$H_0$}
    \vskip 3.24cm ~
    \vskip -5.6cm \hskip 6.8cm {\footnotesize $H_0$}
    \vskip 4.74cm ~
    \vskip -7.1cm \hskip 2.2cm {\footnotesize \emph{CMB}+\emph{CL}}
    \vskip 6.24cm ~
    \vskip -6.3cm \hskip 4.1cm {\footnotesize \emph{CMB}+\emph{BAO}}
    \vskip 5.44cm ~
  \end{minipage}
  \begin{minipage}{0.48\linewidth}
    \smfigure{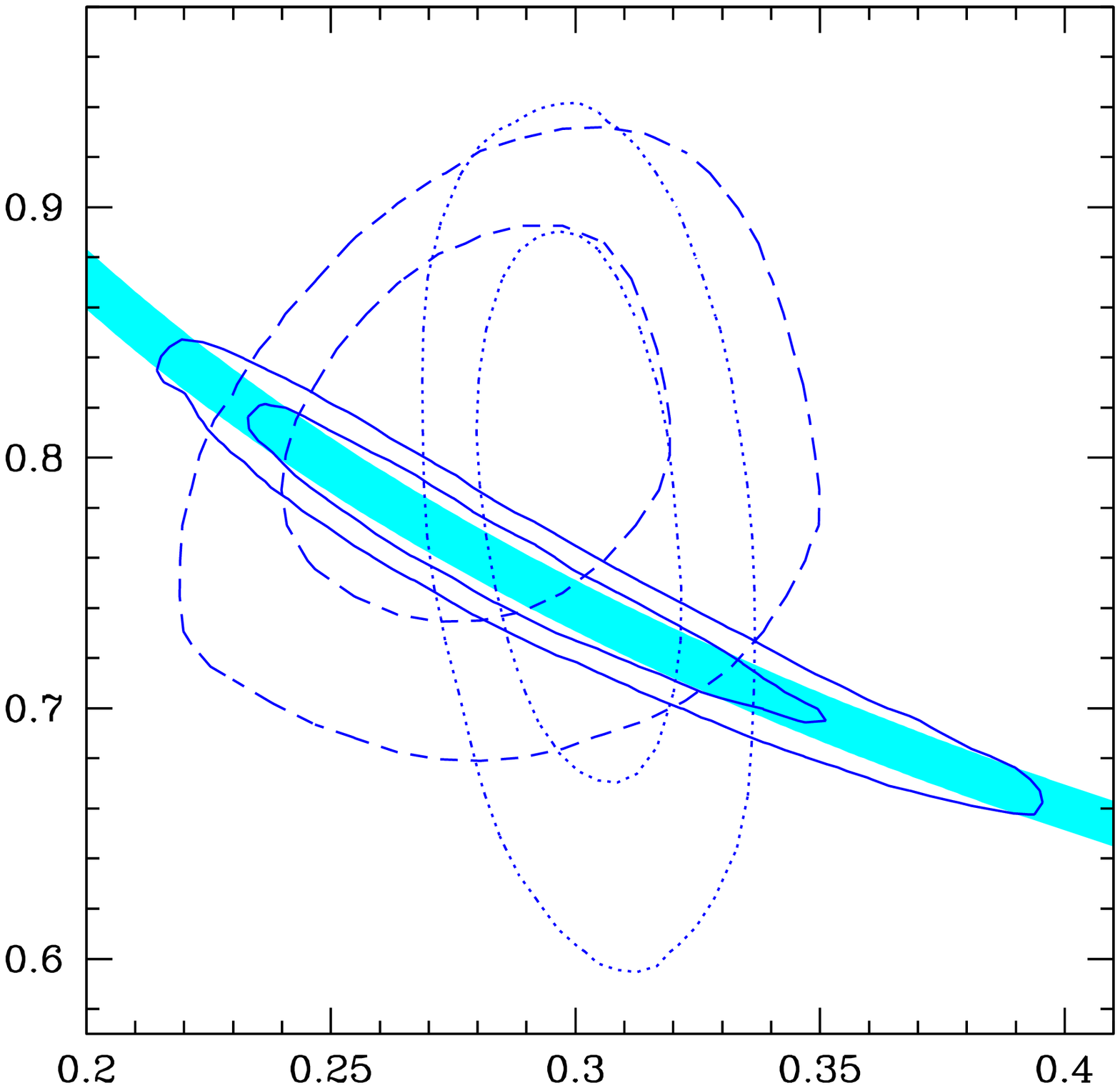}{$\Omega_m$}{$\sigma_8$}
    \vskip -7.9cm \hskip 1.25cm {\footnotesize
      $\Lambda$CDM+\neff+$\Sigma m_\nu$}
    \vskip 7.04cm ~
    \vskip -2.cm \hskip 5.1cm {\footnotesize \emph{CMB}+\emph{BAO}}
    \vskip 1.14cm ~
    \vskip -3.5cm \hskip 5.9cm {\footnotesize \emph{CMB}+\emph{CL}}
    \vskip 2.64cm ~
    \vskip -2.9cm \hskip 1.7cm {\footnotesize \emph{CMB}+$H_0$}
    \vskip 2.04cm ~
    \vskip -6cm \hskip 1.5cm {\footnotesize \emph{CL}}
    \vskip 5.14cm ~
  \end{minipage}
  \caption{The constraints on $\Omega_m$, $\sigma_8$ and $H_0$ in
    $\Lambda$CDM+\neff+$\Sigma m_\nu$ model, obtained using various
    cosmological datasets. Contours are the same as in
    Fig.~\ref{fig:lcdmoms8h0}.}
  \label{fig:lcdmoms8h0nnumnu}
\end{figure*}

\section{$\Lambda$CDM model}

The constraints on the mean matter density in Universe, $\Omega_m$,
the linear density fluctuations amplitude, $\sigma_8$, and Hubble
constant, $H_0$, which are obtained using various cosmological
datasets in standard six-parameter $\Lambda$CDM model
\citep[e.g.,][]{larson11} are shown in Fig.~\ref{fig:lcdmoms8h0}. One
can see that there is a discrepancy between the direct Hubble constant
measurements and the measurement obtained from BAO data in assumption
of $\Lambda$CDM model (see left panel of the Figure). One can also see
that new BAO data produce a discrepancy between the measurements of
$\sigma_8$ from \emph{CMB}+\emph{BAO} and from galaxy cluster mass
function data (see right panel of Fig.~\ref{fig:lcdmoms8h0}).

The discrepancy in Hubble constant measurements appears most
prominently in BAO data from SDSS DR9 (see Fig.~21 and Fig.~30 in
\citealt{anderson12}). This discrepancy was noticed and was discussed
in detail by \cite{anderson12}. In this work BAO data are compared to
the measurement of Hubble constant from \cite{riess11}. When another,
more recent measurement of Hubble constant from \cite{freedman12}, is
taken in account, statistical significance of this discrepancy becomes
more significant. It is this discrepancy, which is shown in the left
panel of Fig.~\ref{fig:lcdmoms8h0}.

\section{Additional neutrino species}

The underestimated measurement of Hubble constant can be obtained from
BAO data in assumption of $\Lambda$CDM cosmological model with
standard number of neutrino species, if in real Universe the extra
component in relativistic energy density is present at early epoch in
addition to photons and three known neutrinos. In this case the
Universe expansion rate is increased during the radiation-dominated
era, and the size of sound horizon is therefore decreased. If one use
$\Lambda$CDM model with standard number of neutrinos, the size of
sound horizon would be overestimated, as compared to its real
size. Since BAO observations give the measurement of the distance in
units of sound horizon, Hubble constant, measured in this way will be
underestimated. Useful discussion on related subjects can be found in
\cite{2004PhRvD..70j3523E} and \cite{hou11}.

Relativistic energy density in early Universe is usually parametrized
using the effective number of neutrino species, \neff:
$$\rho_r=\left[1+\frac{7}{8}\left(\frac{4}{11}\right)^{4/3}\neff\right]\rho_\gamma$$
where $\rho_\gamma$ --- energy density of photons. The constraints on
$\Omega_m$, $\sigma_8$ and $H_0$ in \emph{$\Lambda$CDM} model with
free effective number of neutrino species added are shown in
Fig.~\ref{fig:lcdmoms8h0nnu}. From this Figure one can see that the
discrepancies in Hubble constant measurements are eliminated in this
cosmological model, as expected.

\begin{figure*}
  \centering
  \begin{minipage}{0.48\linewidth}
    \smfigure{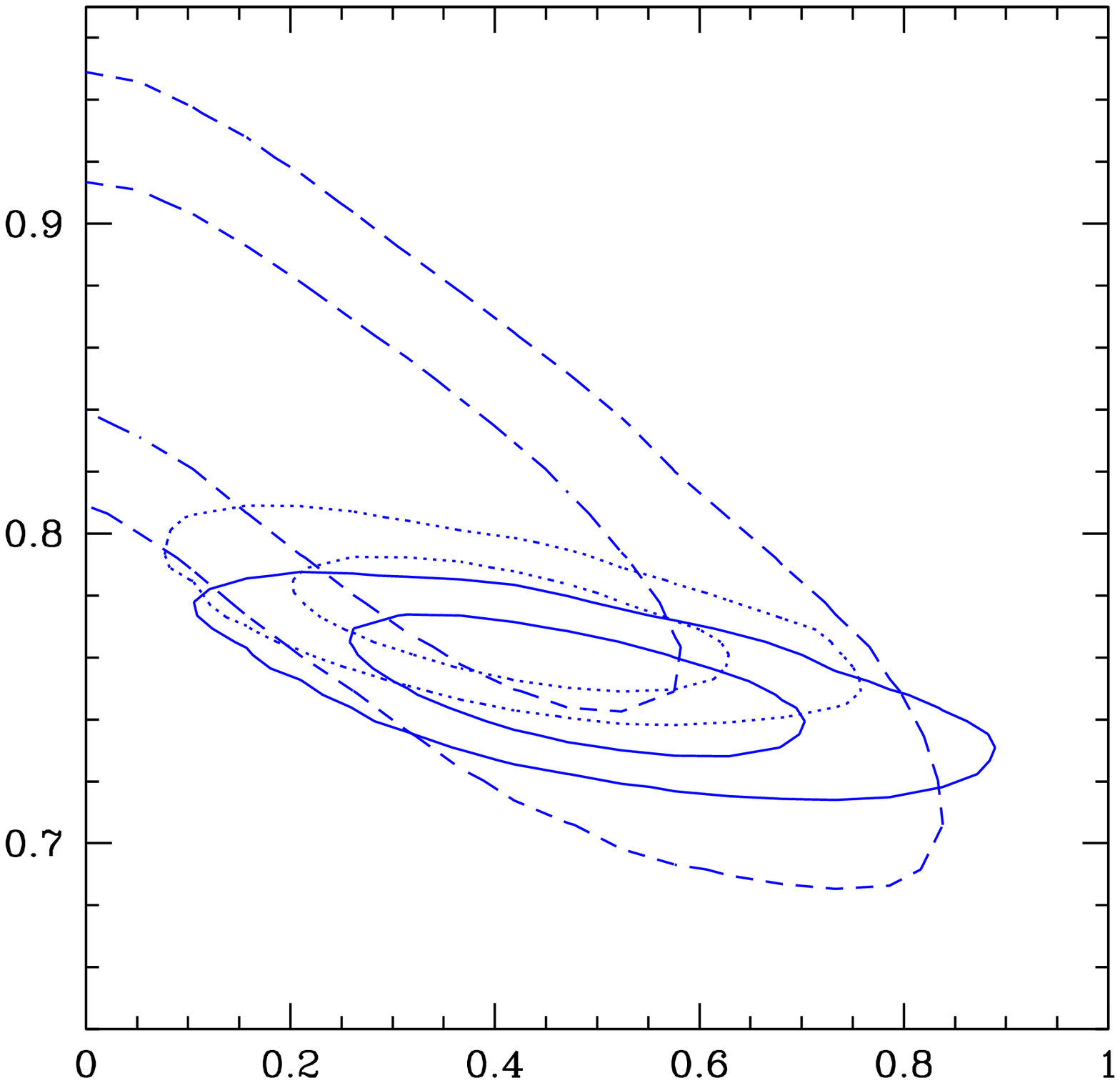}{$\Sigma m_\nu$}{$\sigma_8$}
  \end{minipage}
  \begin{minipage}{0.48\linewidth}
    \smfigure{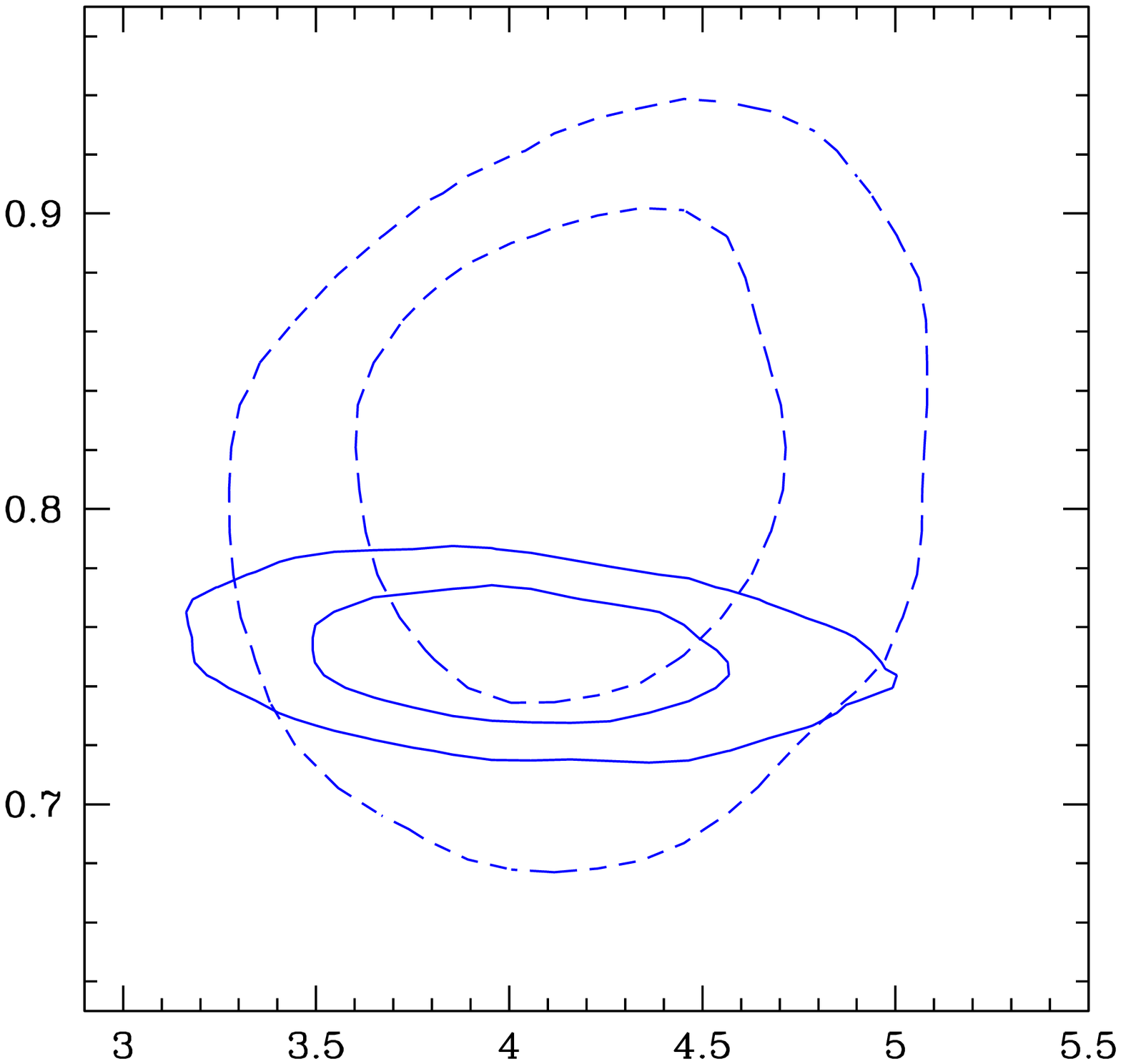}{\neff}{$\sigma_8$}
  \end{minipage}
  \caption{The constraints on $\Sigma m_\nu$ and \neff\ in
    \emph{$\Lambda$CDM}+\neff+$\Sigma m_\nu$ model. The dashed
    contours show the constraints obtained using
    \emph{CMB}+\emph{BAO}+$H_0$ data, solid contours --- the
    constraints with galaxy cluster mass function data taken in
    account, \emph{CMB}+\emph{BAO}+$H_0$+\emph{CL}. In the left panel
    the contours obtained with the same data assuming that the masses
    of clusters in \cite{av09a} are underestimated by 9\% are also
    shown with dotted lines.}
  \label{fig:s8mnunnu}
\end{figure*}

\section{Additional neutrino species and non-zero total neutrino mass}

In Fig.~\ref{fig:lcdmoms8h0nnu} one can also see that even if
additional neutrino species are introduced into $\Lambda$CDM model,
not all the discrepancies are eliminated. From the right panel of this
Figure one can see that the combined data on CMB power spectrum, BAO
and Hubble constant measurements suggest significantly higher value of
$\sigma_8$, as compared to the cluster mass function data at the same
value of $\Omega_m$.

The measurement of $\sigma_8$ from \emph{CMB}+\emph{BAO}+$H_0$ data is
inferred from the amplitude of CMB temperature fluctuation, which
contain the information on density fluctuation amplitude at high
redshifts, $z\approx1000$. On the other hand, the measurement of
$\sigma_8$ from the galaxy cluster mass function data reflects the
value of this quantity at recent epoch. Therefore, this discrepancy in
$\sigma_8$ measurements can be interpreted as a result of a
suppression of density fluctuations due to non-zero total neutrino
mass.

\begin{figure}[t]
  \centering \smfigure{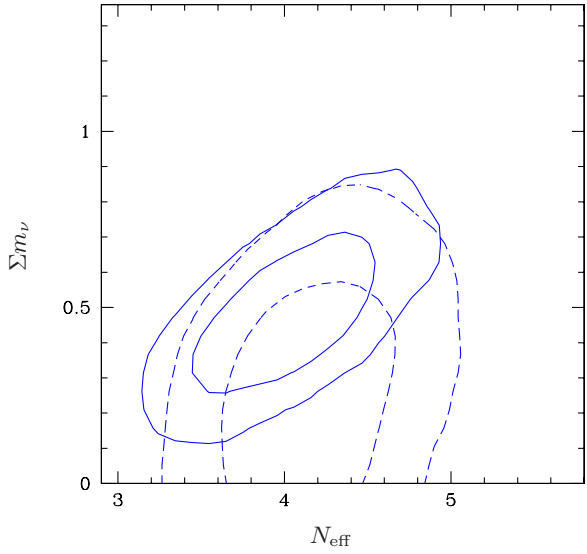}{\neff}{$\Sigma m_\nu$}
  \caption{The constraints on $\Sigma m_\nu$ and \neff\ in
    \emph{$\Lambda$CDM}+\neff+$\Sigma m_\nu$ model. The dashed
    contours show the constraints obtained using
    \emph{CMB}+\emph{BAO}+$H_0$ data, solid contours --- the
    constraints with galaxy cluster mass function data taken in
    account, \emph{CMB}+\emph{BAO}+$H_0$+\emph{CL}.}
  \label{fig:mnunnu}
\end{figure}

The constraints on $\Omega_m$, $\sigma_8$ and $H_0$ in
\emph{$\Lambda$CDM}+\neff+$\Sigma m_\nu$ model are shown in
Fig.~\ref{fig:lcdmoms8h0nnumnu}. One can see that in this case all
cosmological datasets are completely consistent with each other. The
consistency is achieved at the expense of the detection of larger then
standard effective number of neutrino species and non-zero total
neutrino mass. The constraints on total neutrino mass and effective
number of neutrino species obtained with all the data combined are
shown in Fig.~\ref{fig:s8mnunnu} and \ref{fig:mnunnu}.

If all considered cosmological data are used
(\emph{CMB}+\emph{BAO}+$H_0$+\emph{CL}), the change of $\chi^2$ when
two parameters, $\Sigma m_\nu$ and \neff, are introduced into the
cosmological model turns out to be $\Delta\chi^2=13.0$, which
corresponds to $\approx 3.2\sigma$ significance. In this case the
following measurements of effective number of neutrino species and
total neutrino mass are obtained: $\neff=4.03\pm0.36$, $\Sigma
m_\nu=0.49\pm0.17$~eV. Here and everywhere below the systematic
uncertainties of galaxy cluster mass function measurements from
\cite{av09b} are included in the error of total neutrino mass
measurement. Note, that the measurement of effective number of
neutrino species do not depend on cluster mass function data (see
right panel in Fig.~\ref{fig:s8mnunnu}). For the case, when the
measurement of Hubble constant from \cite{riess11} only is used, the
change of $\chi^2$ when two parameters are added to cosmological model
is $\Delta\chi^2=11.9$, which corresponds to $\approx 3.0\sigma$
significance. In this case the following measurements are obtained:
$\neff=3.89\pm0.39$, $\Sigma m_\nu=0.47\pm0.16$~eV.

\begin{figure}[t]
  \centering \smfigure{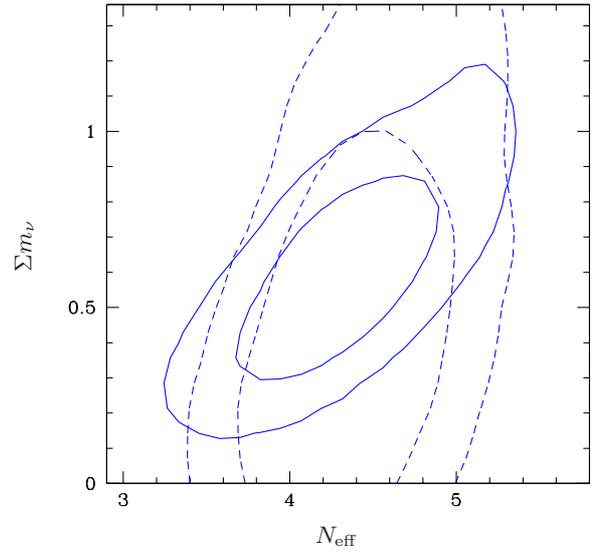}{\neff}{$\Sigma m_\nu$}
  \caption{The constraints on $\Sigma m_\nu$ and \neff\ in
    \emph{$\Lambda$CDM}+\neff+$\Sigma m_\nu$ model in the case when
    only one neutrino type is massive. The contours are the same as in
    Fig.~\ref{fig:mnunnu}}
  \label{fig:mnunnum1}
\end{figure}

When these constraints are obtained, the total neutrino mass is
assumed to be distributed in equal parts among three neutrino types
(this is the default setting in used version of \texttt{CosmoMC}
software). Different distribution of neutrino mass can produce notable
changes in total neutrino mass constraints since, with the same total
neutrino mass, more massive neutrinos become non-relativistic at
earlier time. Correspondent differences in total neutrino mass
constraints are indeed notable when current observational data are
used \citep[e.g.,][]{bv12}. For this reason we also obtained
constraints in the case when only one neutrino type is massive. These
constraints are shown in Fig.~\ref{fig:mnunnum1}. In this case the
upper limit for total neutrino mass is somewhat relaxed, while the
statistical significance of the measurement of non-zero neutrino mass
remains to be approximately the same.

In this case the change of $\chi^2$ when two parameters are added to
the model is $\Delta\chi^2=14.6$, which corresponds to $\approx
3.4\sigma$ significance. In this case the following measurements are
obtained: $\neff=4.32\pm0.39$, $\Sigma m_\nu=0.60\pm0.22$~eV. If the
Hubble constant measurement from \cite{riess11} only is used, we get
$\Delta\chi^2=12.8$ which corresponds to $\approx 3.1\sigma$
significance, and in this case the measurements: $\neff=4.20\pm0.46$,
$\Sigma m_\nu=0.58\pm0.23$~eV are obtained.

\section{Discussion}

As it is shown in Fig.~\ref{fig:lcdmoms8h0}, all the discrepancies in
cosmological data in $\Lambda$CDM model with zero neutrino mass and
standard number of neutrino species appear mainly due to the usage of
new baryon acoustics oscillations data. In fact just these data
require the introduction of additional parameters into $\Lambda$CDM
model, like non-zero total neutrino mass and larger than standard
number of neutrino species. If these BAO data are excluded from the
consideration, all other data remain to be consistent with standard
$\Lambda$CDM model (see Fig.~\ref{fig:lcdmoms8h0}).

We note that in addition to the data used in our work, the other BAO
observations exist where the lower value of Hubble constant is
obtained in $\Lambda$CDM model, as compared to direct Hubble constant
measurements. For example, similar discrepancy was found in the
measurement of angular diameter distance from BAO observed using
photometric redshifts of luminous red galaxies \citep{seo12}. The data
on BAO observations in the transmitted flux fraction in the Ly$\alpha$
forest of high redshift quasars obtained using SDSS DR9 data
\citep{busca12} published recently, also give underestimated Hubble
constant in $\Lambda$CDM model \citep[see.\ Fig.\ 21
in][]{busca12}. Note, that these BAO data are obtained using
significantly different methods, as compared to BAO measurements in
large spectroscopic galaxy surveys used in our work.

As it was discussed above, this discrepancy was noticed earlier, and
it was also found that it can be eliminated if extra energy density of
relativistic matter is present in early Universe
\citep{mehta12,anderson12,freedman12}. Independent indications for the
possibility of larger than standard value of \neff, were also obtained
earlier using the data on CMB anisotropy at smaller angular scales
(\citealt{dunkley11,keisler11}, see also discussion in
\citealt{hamann11,joudaki12,2013ApJ...763...89R}). In this case the
additional radiation energy density in early Universe is observed
using different physical effect, by the measurement of reduced power
in CMB power spectrum damping tail due to larger photon diffusion
angular scale. These measurements are degenerate with primordial
helium abundance and running spectral index (see details, e.g., in
\citealt{dunkley11} and \citealt{keisler11}). The data from
\cite{keisler11} are used in our work as well, and they also have some
influence on our \neff\ measurement, however, this influence is not
dominant. As it was discussed above, the measurement of \neff,
presented in our work, is based mainly on the new data on BAO
observations compared to the direct $H_0$ measurements.

From Fig.~\ref{fig:lcdmoms8h0} one can also see that current data on
direct Hubble constant measurements and cluster mass function are
consistent with each other in standard $\Lambda$CDM model. In order to
bring these data in accordance to recent BAO observations one need to
decrease the measured value of Hubble constant by 7--10\% and to
\emph{simultaneously} increase the value of $\sigma_8$ measured at the
same $\Omega_m$ also by 7--10\%.

In our work we use two independent measurement of Hubble constant
which are in good agreement with each other
\citep{riess11,freedman12}. Each of them is accurate to approximately
3\%, including systematic errors. These are the data which give the
measurement of larger than standard neutrino species when are compared
to BAO observations.

The data on galaxy cluster mass function in their turn, when compared
to the data on BAO observations, give the measurement of non-zero
total neutrino mass. These constraints are affected by systematic
errors of cluster mass function measurements. However, these data
would be consistent with zero total neutrino mass only if the masses
of clusters in \cite{av09a} were underestimated by approximately 30\%.
One can see this from the left panel in Fig.~\ref{fig:s8mnunnu}, where
the dotted lines show the contours for all considered data
(\emph{CMB}+\emph{BAO}+$H_0$+\emph{CL}), in case if cluster masses in
\cite{av09a} are underestimated by 9\%, the adopted systematic error
for cluster mass scale calibration.

Systematic errors in determination of cluster mass scale is one of the
main systematic uncertainties in cluster mass function
measurements. In \cite{av09b} the measurements of cluster masses were
based on the temperature and mass of hot intracluster gas and were
calibrated using hydrostatic mass measurements. Systematic error of
cluster mass scale calibration was estimated as $\delta M/M \approx
0.09$, using the comparison of hydrostatic cluster masses with those
based on weak lensing measurements, taken from \cite{hoekstra07} and
\cite{zhang08}. More recent works, where the additional weak lensing
data are presented, are in general agreement with approximately 10\%
systematic error for cluster masses measured using the data of X-ray
observations \citep{israel10,israel12,mahdavi12,applegate12}.

\section{Conclusions}

In our work we show that the discrepancies between different
cosmological datasets in the determination of Hubble constant and in
measurements of density fluctuation amplitude in assumption of
standard $\Lambda$CDM cosmological model, can be eliminated if the
additional neutrino species and non-zero total neutrino mass are
introduced into the cosmological model. In this case, the discrepancy
in the determination of distance scale between baryon acoustic
oscillations data and direct Hubble constant measurement is eliminated
by the assumption of larger than standard effective number of neutrino
species, which was already noticed earlier \citep[see,
e.g.,][]{mehta12,anderson12,freedman12,seo12}. The remaining
discrepancy in $\sigma_8$ measurements between combined CMB, BAO and
Hubble constant data and the data on galaxy cluster mass function is
eliminated by the assumption of non-zero neutrino mass.

The change of $\chi^2$ when two parameters, \neff\ and $\Sigma m_\nu$,
are introduced into the cosmological model corresponds to
$\approx3\sigma$ significance level. The model with approximately one
additional neutrino type, $\neff\approx 4$, and with non-zero total
neutrino mass $\Sigma m_\nu\approx 0.5$~эВ provide the best fit to the
data. In model with only one massive type of neutrino the upper limits
on neutrino mass are slightly relaxed.

We emphasize, that $\Lambda$CDM model with standard number of neutrino
species and zero neutrino mass appears to be no longer consistent with
cosmological data due to the usage of the data on baryon acoustic
oscillations, published recently. In future these results may be
independently confirmed using significantly improved data on CMB
anisotropy at small angular scales. In very near future the
measurements made with Planck space observatory should be published
which will probably clarify this issue.

\begin{figure}[t]
  \centering \smfigure{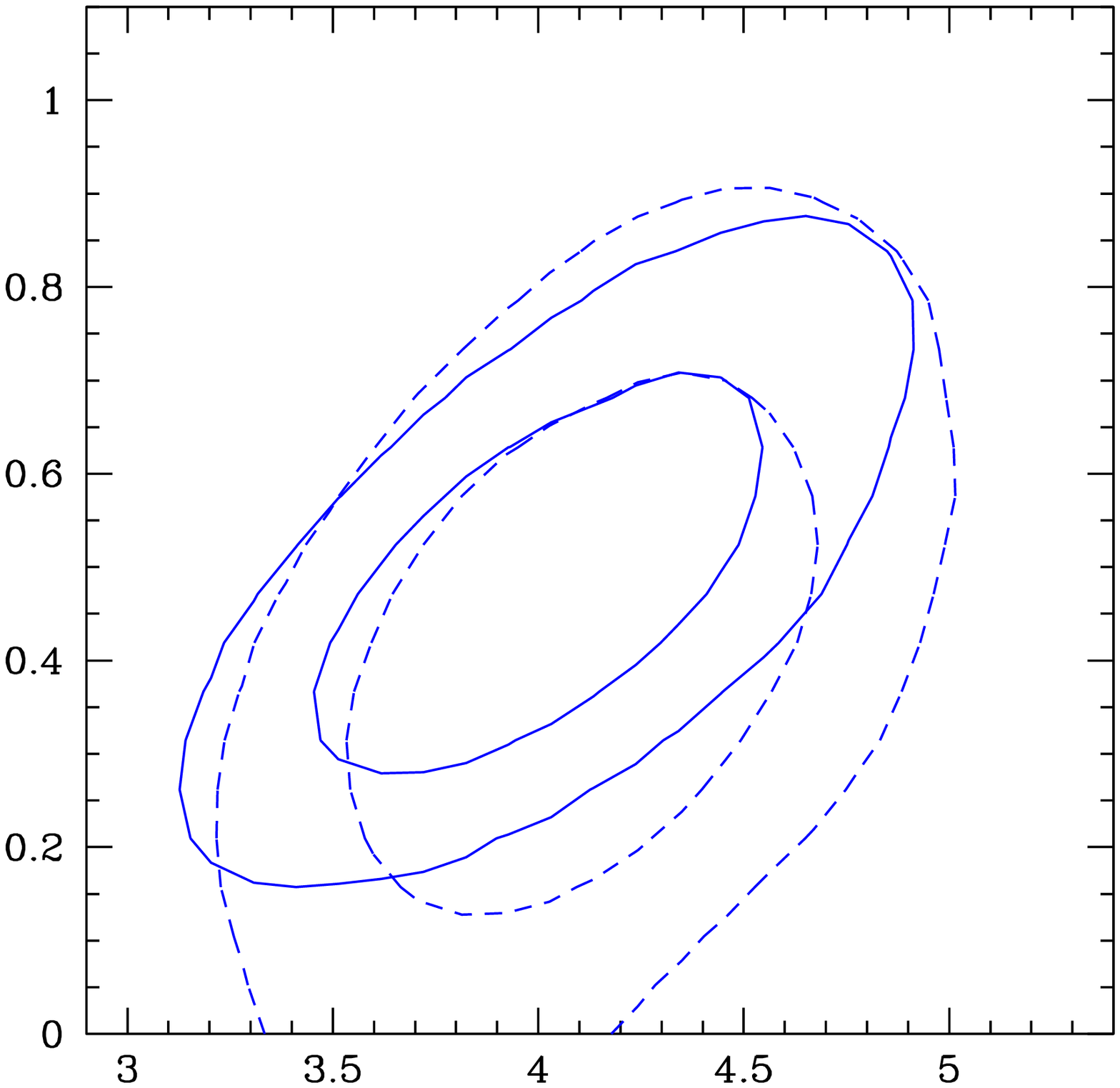}{\neff}{$\Sigma m_\nu$}
  \caption{The constraints on $\Sigma m_\nu$ and \neff\ in
    \emph{$\Lambda$CDM}+\neff+$\Sigma m_\nu$ model, obtained using
    \emph{WMAP9} data and also with the data on CMB lensing potential
    included.  The dashed contours show the constraints obtained using
    \emph{WMAP9}+SPT+CMBLens+\emph{BAO}+$H_0$ data, solid contours ---
    the constraints with galaxy cluster mass function data taken in
    account, \emph{WMAP9}+SPT+CMBLens+\emph{BAO}+$H_0$+\emph{CL}.}
  \label{fig:mnunnuwmap9lens}
\end{figure}

\begin{figure}[t]
  \centering \smfigure{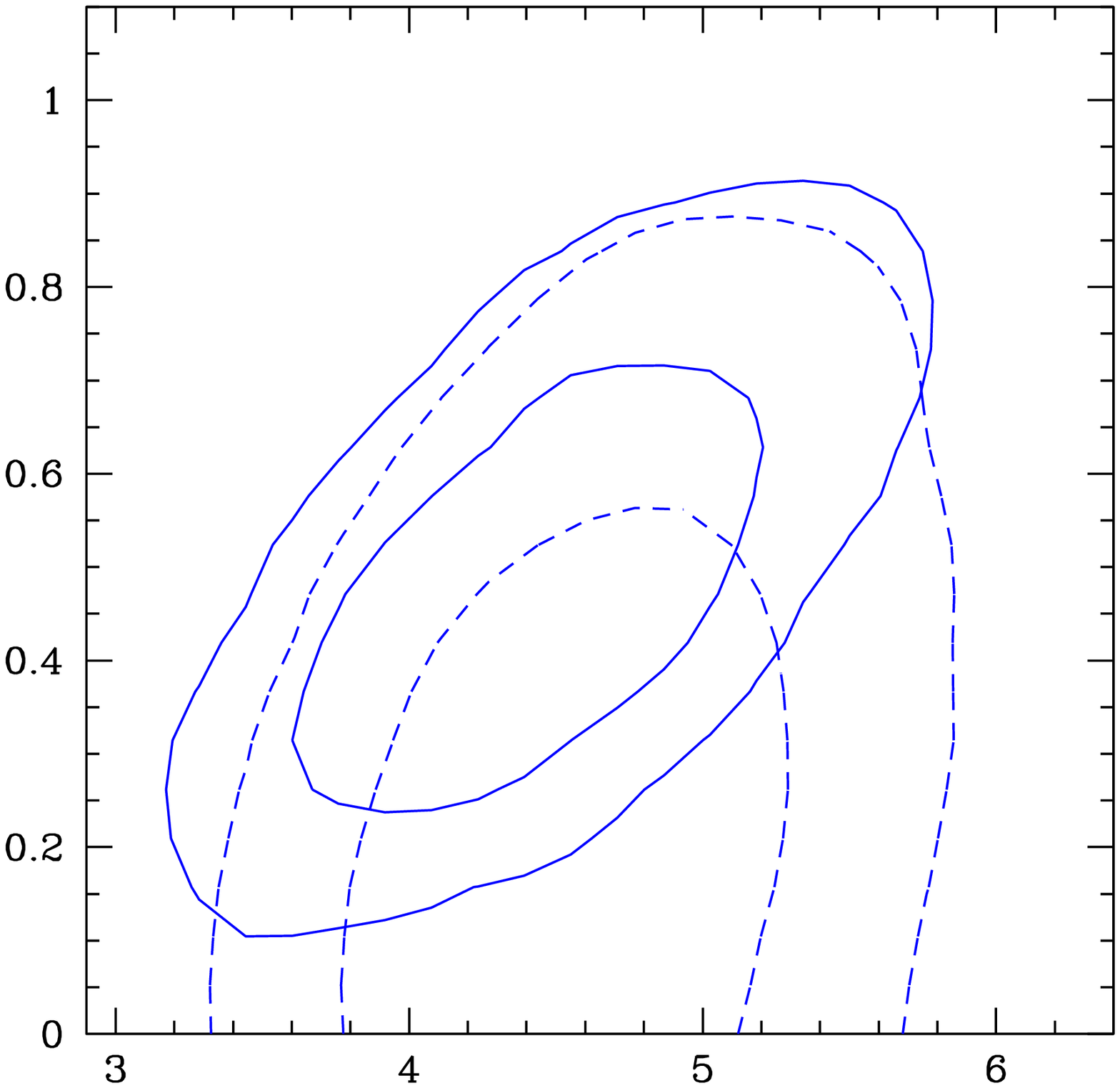}{\neff}{$\Sigma m_\nu$}
  \caption{The constraints on $\Sigma m_\nu$ and \neff\ in
    \emph{$\Lambda$CDM}+\neff+$\Sigma m_\nu$ model using all
    considered data, excluding the data from \emph{SPT}. The dashed
    contours show the constraints obtained using
    \emph{WMAP7}+\emph{BAO}+$H_0$ data, solid contours ---
    the constraints with galaxy cluster mass function data taken in
    account, \emph{WMAP7}+\emph{BAO}+$H_0$+\emph{CL}.}
  \label{fig:mnunnunospt}
\end{figure}

It is known that the existence of additional light neutrino species,
which could produce oscillations with $\Delta m^2\sim 1$~eV$^2$, was
suggested to explain the results of the experiments of neutrino
oscillations searches at short baselines --- \emph{LSDN}
\citep{aguilar01}, \emph{MiniBooNE} \citep{aguilararevalo10}, reactor
antineutrino anomaly \citep{mueller11,mention11}, and gallium anomaly
\citep{2006PhRvC..73d5805A,2010PhLB..685...47K,2011PhRvC..83f5504G}. In
order to explain the results of these experiments the possibility of
the existence of one or two light sterile neutrino was discussed
\citep[see details in, e.g.,][]{abazajian12}.

These neutrinos can be constrained from the cosmological data as
well. The constraints presented in our work are consistent with the
existence of one additional type of neutrino. Since in this case
$\Delta m \sim m$, the masses of all other neutrinos should be
significantly lower, and to constrain these neutrinos one should use
the model with only one massive type of neutrino. These constraints
were obtained above and they are as follows: $\neff=4.32\pm0.39$,
$\Sigma m_\nu=0.60\pm0.22$~eV.

\paragraph{Note} 

When this paper was already submitted for publication, few papers were
published, where some new constraints on number of neutrino species
and total neutrino mass based on new cosmological data were presented.
They are the papers on cosmological constraints obtained from nine
year \emph{WMAP} data \citep{hinshaw12} and from new measurements of
CMB anisotropy at high multipoles using the data from South Pole
Telescope \citep[\emph{SPT,}][]{hou12} and Atacama Cosmology Telescope
\citep[\emph{ACT,}][]{sievers13}.

In their paper on nine year \emph{WMAP} data, \cite{hinshaw12},
obtained $\neff=3.84\pm0.40$ in model with free \neff\ and zero total
neutrino mass, in agreement with our constraints. Note, that in this
work CMB data include also the data on power spectrum of CMB
gravitational lensing potential from \cite{das11} and
\cite{vanengelen12}, which contain the information on the density
fluctuation amplitude at redshifts approximately $0.5<z<5$. If the
data on density fluctuations amplitude at recent epoch are added to
other current cosmological data in $\Lambda$CDM+\neff\ model with zero
neutrino mass, the lower value of \neff\ is measured due to the
degeneracy between \neff\ and $\Sigma m_\nu$, which appears in this
case (\citealt{bv12}, see also, e.g., contours for \neff--$\Sigma
m_\nu$ in Figures above).

The effect of these new data on the constraints presented above is
shown in Fig.~\ref{fig:mnunnuwmap9lens}. In this Figure the
constraints on number of neutrino species and total neutrino mass
obtained using the latest 9-year WMAP data \citep[WMAP9,][]{hinshaw12}
and also the data on CMB lensing power spectrum amplitude
(\mbox{CMBLens}), from \cite{das11} and \cite{vanengelen12}, which was
taken in account as a gaussian prior, $C_{400}^{\kappa \kappa} = (3.17
\pm 0.45)\times 10^{-8}$, as it was done by \cite{calabrese13}, are
shown. The new data only slightly change the results presented
above. In this case the significance of the inclusion of two
parameters, \neff\ and $\Sigma m_\nu$ in the model is
$\approx3.3\sigma$, and we obtained the following measurements:
$\neff=4.02\pm0.35$ and $\Sigma m_\nu=0.50\pm0.15$~eV. If only one
type of neutrino is massive, the correspondent significance is
approximately the same and we obtain: $\neff=4.12\pm0.36$ and $\Sigma
m_\nu=0.58\pm0.21$~eV. Note, that when CMB lensing power spectrum data
are used, non-zero neutrino mass is obtained with lower significance
even without clusters mass function data (dashed contours in
Fig.~\ref{fig:mnunnuwmap9lens}).

The constraints based on new data from South Pole Telescope
\citep{hou12} in $\Lambda$CDM+\neff+$\Sigma m_\nu$ model are generally
consistent with the results, obtained in our work. From Fig.~18 of
\cite{hou12} one can see that, similarly to our work, the measurement
of larger than standard effective number of neutrino species is based
mainly on the new data on BAO observations (see also the discussion in
\S 10 of \citealt{hou12}). Also, as in our work, the measurement of
non-zero neutrino mass comes from the data on galaxy clusters mass
function. Therefore, as compared to our work, \cite{hou12} used
approximately similar cosmological data and obtained approximately
similar constraints on neutrino properties, as expected.

In their work, based on the new data of Atacama Cosmology Telescope
(\emph{ACT}), \cite{sievers13} did not find additional neutrino types
when both \emph{$\Lambda$CDM}+\neff\ and
\emph{$\Lambda$CDM}+\neff+$\Sigma m_\nu$ models are considered. Note,
that not complete BAO dataset is used in this work (only the data from
SDSS DR7, \citealt{percival10}, \emph{6dF}, \citealt{beutler11}, and
SDSS DR9, \citealt{anderson12}). Also, as in \cite{hinshaw12} and
\cite{hou12}, the $H_0$ measurement from \cite{freedman12} is not
used. Apparently, in the new data from \emph{ACT} there are some
statistically insignificant discrepancies with \emph{SPT} data which
bring the measurement of \neff\ to some lower value. However, as it
was discussed above, in our work the measurements of larger than
standard number of neutrinos and non-zero neutrino mass are based
mainly on recent BAO data and do not depend strongly on the data on
high-$l$ CMB measurements.

In order to better show this, in Fig.~\ref{fig:mnunnunospt} the
constraints on \neff\ and $\Sigma m_\nu$ are given, which were
obtained using all the data considered in our work, excluding the
high-$l$ CMB data from \emph{SPT} (i.e., using the
\emph{WMAP7}+\emph{BAO}+$H_0$+\emph{CL} dataset). There is no much
change in the significance of the detection of $\neff>3$ and $\Sigma
m_\nu>0$, the main effect of the exclusion of \emph{SPT} data is the
relaxed upper limit for \neff. In this case, the change of $\chi^2$
when two parameters, \neff\ and $\Sigma m_\nu$, are introduced into
the model is $\Delta\chi^2=11.1$, which corresponds to $\approx
2.9\sigma$ significance, and the following measurements are obtained:
$\neff=4.44\pm0.50$, $\Sigma m_\nu=0.49\pm0.18$~eV. In the case, when
only one neutrino type is massive, $\Delta\chi^2$ have the same value,
and the following measurements are obtained:$\neff=4.55\pm0.50$,
$\Sigma m_\nu=0.54\pm0.24$~eV.

\acknowledgements 

Author is grateful to A.~A.~Vikhlinin for critical discussion and
useful comments. In this work the results of computations made with
MVS-100K supercomputer of Joint Supercomputer Center of the Russian
Academy of Sciences (JSCC RAS) were used. The work is supported by
Russian Foundation for Basic Research, grants 10-02-01442,
11-02-12271-ofi-m, 12-02-01358, the Program for Support of Leading
Scientific Schools of the Russian Federation (NSh-5603.2012.2), and
the Programs of the Russian Academy of Sciences P-21/1 and OPhN-17.

\end{document}